**ORBITS OF DUST EJECTA FROM CERES.** R. C. Nazzario and T. W. Hyde, Center for Astrophysics, Space Physics and Engineering Research, Baylor University, Waco, TX 76798-7316, USA, phone: 254-710-2511 (email: Truell_Hyde@Baylor.edu).

**Introduction:** Many asteroids show visual evidence of cratering with the apparent size of the craters revealing a large range of impactor size [1-3]. Studies primarily concentrating on disruption or fragmentation of the asteroid [4] or investigating stable orbits for both uncharged particles orbiting the asteroid [5,6] and charged particles levitated by electrostatic means from the asteroid's surface [7] have been conducted. However, even during non-catastrophic impacts, ejecta of many different sizes will be produced. Of particular interest are the larger dust particles (from 1 µm to 1000 µm in radius) which suffer one of three possible fates. First, they can be re-accreted by the asteroid. Secondly, they can enter into orbit around the asteroid. Finally, they can escape the gravitational pull of the asteroid altogether. The second scenario is of greatest interest to planners of asteroid rendezvous spacecraft, as spacecraft (and their payload instruments) are susceptible to damage during encounters with such dust [8]. Additionally, for maximum efficiency dust detection systems must be specifically configured for a particular size and speed distribution of the dust. Although dust particles have not yet been directly observed around asteroids, in-situ data are extremely limited since only one such environment has been investigated at close range [3]. This study concentrates on the question of whether dust ejecta can enter into stable orbits around an asteroid and if so the size and speed distribution of such dust.

**Model:** It will be assumed that an impact on an asteroid will eject dust particles in much the same manner as previously observed for the moons of Jupiter and Earth [9]. After ejection, the orbital parameters for each grain are calculated taking into account the asteroid's gravitational force, the solar radiation pressure force, the solar gravitational force, and the interaction of the dust particle with the Solar magnetic field as given in Eq. (1) below.

$$\vec{a} = -\frac{GM_A}{r^2}\hat{r} - \frac{\beta GM_{sun}}{R^2}\hat{R} + \frac{3\varepsilon_o \phi}{\rho_d s^2}\vec{v}\times\vec{B} + M_{sun}G\left[-\frac{\vec{r}}{R^3} + \vec{\rho}\left(\frac{1}{R^3} - \frac{1}{\rho^3}\right)\right] \quad (1)$$

In Eq. (1), the first and fourth terms deal with the gravitational acceleration due to the asteroid and Sun respectively where G is the gravitational constant, $M_A$ is the mass of the asteroid, r is the distance from the center of the asteroid to the position of the particle, $\hat{r}$ is the unit vector from the asteroid to the particle, $M_{Sun}$ is the mass of the Sun, R is the distance from the Sun to the particle, and $\hat{R}$ is the unit vector from the Sun to the dust particle. In the last term, ρ is the distance and $\hat{\rho}$ is the unit vector from the Sun to the asteroid [10].

The second term describes the radiation pressure force due to the solar radiation incident upon the particle with β defined as the ratio of the solar radiation pressure force to the solar gravitational force [11]. β is calculated using

$$\beta = \frac{.6Q}{a\rho_d} \quad (2)$$

where Q is the radiation pressure efficiency (taken to be 1.0 for particles with radii larger than 1 µm), a is the radius of the dust particle, and $\rho_d$ is its density. The third term in equation (1) takes into account the interaction of the magnetic field on a charged particle where ϕ is the potential of the grain, $\vec{B}$ is the solar magnetic field, $\vec{v}$ is the velocity of the dust grain relative to the magnetic field and $\varepsilon_o$ is the permittivity of free space. Recent studies by Kimura and Mann [12] indicate that particles exposed to a solar system environment outside of 1 AU will be charged to a potential of —3.0 V. Since asteroids do not provide significant protection from the interplanetary environment, this potential is assumed for all the dust particles examined in this study.

The asteroid Ceres was chosen since its mass makes it more likely to confine particles to a stable orbit, and its almost spherical shape reduces the complexity of the simulation. The dust was ejected from a randomly chosen point on the surface of Ceres in a direction parallel to the surface and with speeds ranging from 300 to 600 m/s. This distribution corresponds with experimental hypervelocity impact data showing that such ejecta are created during low impact velocities [13]. A total of 100 particles were launched at each speed with β values between 0.001 and 0.1 and a step size of 0.001. This β regime corresponds to particle sizes of 220 µm to 2.2 µm respectively, assuming the density of the dust to be 2.4 g/cm$^3$, which is the bulk density of Ceres [14].

**Results:** After ejection, dust particle orbits were tracked for one year. Dust particles with ejection speeds between 430 and 580 m/s entered into orbit around Ceres. The asteroid accreted grains with ejection speeds less than 430 m/s, while particles with ejection speeds greater than 580 m/s escaped its

## ORBITS OF DUST EJECTA FROM CERES: R. C. Nazzario and T. W. Hyde

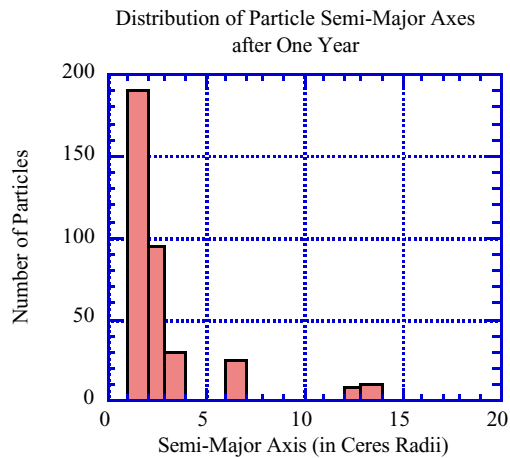

Figure 1. Semi-major axis distribution for particles surviving at least one year.

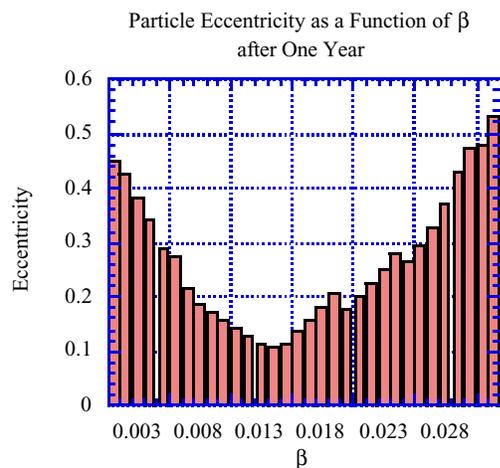

Figure 2. The eccentricity as a function of $\beta$ for particles surviving at least one year.

gravitational sphere of influence altogether. Additionally, all particles ejected at angles less than 85° to the surface normal were removed from the system within a month. Particles entering into long term orbits (i.e. lasting the full year of the simulation) remained within 14 Ceres radii as illustrated by the semi-major axis distribution in Fig. 1. Almost 90% of the particles remained within four Ceres radii of the asteroid. Figure 2 shows the eccentricity as a function of particle size ($\beta$ value) at one year after ejection. It is interesting to note that the minimum eccentricity is for particles with a $\beta$ value approximately equal to 0.013 and not the smallest dust particles as would be expected. Figure 3 illustrates the final positions of the dust one year after ejection and when looking down on the orbital plane of Ceres (the x-y plane). The particles are centered in a plane (from upper left to lower right) with the solar direction as indicated.

**Conclusions:** Of the primary simulations examined (those for particles with ejection speeds between

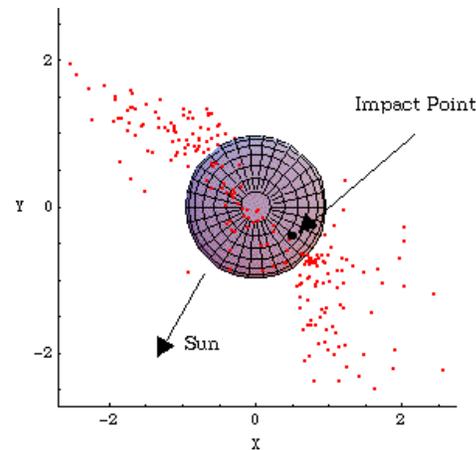

Figure 3. The spatial distribution of surviving dust.

430 m/s and 580 m/s) only 359 particles survived out of the 1600 ejected. All particles smaller than 7.0 µm were removed from the system by the end of one year with 64% of the particles larger than 7.0 µm in radius accreted by Ceres. The orbital speeds for the remaining particles varied from almost 400 m/s for particles with semi-major axes just over one Ceres radii to less than 100 m/s for particles at 14 Ceres Radii. These speeds were size independent and varied primarily with distance from Ceres. Such relatively low speeds even when coupled with the orbital speed of a spacecraft (from 100 to 400 m/s, depending on distance from Ceres) should not pose a significant hazard except for spacecraft in retrograde orbits as compared to the dust grain orbits [15]. The particles were centered on a plane in a torus structure; however, this was likely due to the single ejection point. No ring-like structures were produced within the one year time frame of the simulation. Simulations extending beyond one year in duration are currently in progress.